\documentclass[usenatbib]{mn2e}

\usepackage{amsmath,amssymb, graphics}
\usepackage{pgf,tikz}
\usepackage{bm}
\usepackage{graphicx}
\usepackage{epstopdf}
\usepackage{float}
\usepackage{mathrsfs}

\newcommand{\bl}{\bm {\hat l}}

\newcommand{\rin}{{r_\text{in}}}
\newcommand{\rout}{r_\text{out}}
\newcommand{\Md}{M_\text{d}}

\newcommand{\der}{\text{d}}
\newcommand{\pd}{\partial}

\newcommand{\om}{\omega}

\newcommand{\lam}{\lambda}

\newcommand{\ag}{\alpha}

\newcommand{\Sg}{\Sigma}

\newcommand{\cs}{c_{\rm s}}
\newcommand{\ro}{r_{\rm in}}
\newcommand{\Mb}{M_{\rm b}}
\newcommand{\blb}{\bm {\hat l}_{\rm b}}
\newcommand{\ve}{{\bm e}}
\newcommand{\vr}{{\bm {\hat r}}}
\newcommand{\vphi}{{\bm {\hat \phi}}}
\newcommand{\cO}{\mathcal{O}}

\newcommand{\omb}{\omega_{\rm b}}
\newcommand{\bomb}{\bar \omega_{\rm b}}

\newcommand{\bx}{\bm {\hat x}}
\newcommand{\by}{\bm {\hat  y}}

\newcommand{\im}{\textit{i}}
\newcommand{\Ep}{E_+}
\newcommand{\Em}{E_-}

\newcommand{\bF}{{\bm f}}

\newcommand{\hphi}{\hat {\bm \phi}}
\newcommand{\br}{{\bm r}}
\newcommand{\hr}{\hat {\bm r}}

\newcommand{\bL}{{\bm L}}

\newcommand{\ab}{a_{\rm b}}
\newcommand{\brb}{{\bm r}_{\rm b}}
\newcommand{\bv}{{\bm v}}

\newcommand{\ILK}{I_{\rm LK}}
\newcommand{\domextra}{\omega_{\rm ext}}

\newcommand{\btimes}{{\bm \times}}
\newcommand{\bcdot}{{\bm \cdot}}

\newcommand{\xin}{x_{\rm in}}
\newcommand{\agb}{\alpha_{\rm b}}
\newcommand{\blam}{\bar \lambda}

\newcommand{\be}{\begin{equation}}
\newcommand{\ee}{\end{equation}}

\begin{document}


\title[Hydrodynamical Lidov-Kozai Instability]
{Lidov-Kozai Mechanism in Hydrodynamical Disks: Linear Stability Analysis}


\author[J. J. Zanazzi and Dong Lai]{J. J. Zanazzi$^{1}$\thanks{Email: jjz54@cornell.edu}, and Dong Lai$^{1}$ \\
$^{1}$Cornell Center for Astrophysics, Planetary Science, Department of Astronomy, Cornell University, Ithaca, NY 14853, USA}



\maketitle
\begin{abstract}
Recent SPH simulations by \cite{Martin(2014)} suggest a
circumstellar gaseous disk may exhibit coherent
eccentricity-inclination oscillations due to the tidal forcing of an
inclined binary companion, in a manner that resembles Lidov-Kozai
oscillations in hierarchical triple systems. We carry out linear
stability analysis for the eccentricity growth of circumstellar disks
in binaries, including the effects of gas pressure and viscosity 
and secular (orbital-averaged) tidal force from the inclined
companion. We find that the growth of disk eccentricity
depends on the dimensionless ratio ($S$) between $\cs^2$ (the disk sound
speed squared) and the tidal torque acting on the disk (per unit mass)
from the companion.  For $S\ll 1$, the standard Lidov-Kozai result is
recovered for a thin disk annulus: eccentricity excitation occurs when the
mutual inclination $I$ between the disk and binary lies between $39^\circ$ and
$141^\circ$. As $S$ increases, the inclination window for eccentricity
growth generally becomes narrower. For $S\gtrsim$ a few, eccentricity
growth is suppressed for all inclination angles. Surprisingly,
we find that for $S\sim 1$ and certain disk density/pressure profiles,
eccentricity excitation can occur even when $I$ is much less than $39^\circ$.
\end{abstract}





\section{Introduction}
\label{sec:Intro}

When a test particle orbiting a central mass has a distant binary companion, it can undergo eccentricity and inclination oscillations if the initial inclination $I$ between the orbital planes of the test mass and the binary is sufficiently large.  This is termed Lidov-Kozai (LK) oscillation, and was originally invoked to explain the dynamics of artificial satellites \citep{Lidov(1962)} and asteroids \citep{Kozai(1962)}.  Since then, the LK effect has found a plethora of applications in astrophysics (e.g. \citealt{TremaineYavetz(2014),Naoz(2016)}), such as the formation of the Jovian irregular satellites \citep{Carruba(2002),Nesvorny(2003)}, mergers of massive black hole binaries \citep{Blaes(2002)}, formation of short-period stellar binaries \citep{EggletonKiseleva-Eggleton(2001)} and hot Jupiters \citep{WuMurray(2003),FabryckyTremaine(2007),Petrovich(2015),Anderson(2016)}, 
and Type Ia supernovae from white dwarf binary mergers \citep{KatzDong(2012)}.

The simplest LK oscillation involves only the quadrupole potential from the companion. It has been recognized that the high-order perturbation (e.g., \citealt{Ford(2000),Naoz(2011),Katz(2011)})  and short-range forces (e.g., \citealt{Holman(1997),WuMurray(2003),Liu(2015)}) can significantly influence the LK oscillation dynamics. Thus, one may expect that any eccentricity/inclination oscillations of a gaseous disk inside a stellar binary, if occur at all, may be modified or suppressed by hydrodynamic forces.

Recently \cite{Martin(2014)} used SPH simulations to show that LK oscillations may be excited in circumstellar disks with distant, inclined binary companions (see also \citealt{Fu(2015a)}).  \cite{Fu(2015b)} showed that these disk oscillations can be suppressed by the disk self-gravity when the disk mass is sufficiently large (\citealt{Batygin(2011)}; see discussion in Sec. 4).  If real, this may have interesting astrophysical implications due to the ubiquity of misaligned circumstellar accretion disks in binary systems.

In this paper we use linear theory of eccentric disks\citep{GoodchildOgilvie(2006),Ogilvie(2008),TeyssandierOgilvie(2016)} to study the possibility of coherent LK oscillations of circumstellar disks in binaries.  Section~\ref{sec:formalism} gives the set-up and formalism of this work.  Section~\ref{sec:results} contains our results.  Section~\ref{sec:discussion} presents the summary and discussion of our work.

\section{Setup and Formalism}
\label{sec:formalism}

Consider a circumstellar disk around a host star of mass $M$.  The disk has an inner radius $r = \ro$, outer radius $r = \rout$, and surface density $\Sg = \Sg(r)$.  The disk warp and eccentricity are specified by the unit angular momentum vector $\bl=\bl(r,t)$ and eccentricity vector $\ve=\ve(r,t)$.  We take the disk to be nearly circular, so $e \ll 1$ everywhere.  We adopt a locally isothermal equation of state, so that the height-integrated pressure at any location in the disk is given by $P = \cs^2 \Sg$, where $\cs = \cs(r)$ is the sound speed.  For a thin disk with mass much less than $M$, the orbital frequency of the disk is given by $n(r) \simeq \sqrt{GM/r^3}$.  The host star has a distant external binary companion with semimajor axis $\ab \gtrsim 3\rout$\footnote
{
The upper bound on the outer disk radius is set by tidal truncation \citep{ArtymowiczLubow(1994),MirandaLai(2015)} 
}, mass $\Mb$, and orbital angular momentum unit vector $\blb$.  We take the binary's orbit to be circular.  Because the angular momentum of the binary is much larger than that of the circumstellar disk, we take $\blb$ to be fixed in time.



The gravitational force of the binary companion drives the eccentricity and angular momentum unit vectors of disk annuli according to \citep{Tremaine(2009),TremaineYavetz(2014)}
\begin{align}
\left( \frac{\pd \bl}{\pd t} \right)_{\rm bin} &= \omb (\bl \bcdot \blb) \bl \btimes \blb + \cO(e^2)
\label{eq:dldtb} \\
\left (\frac{\pd \ve}{\pd t} \right)_{\rm bin} &= \omb \Big[ (\bl \bcdot \blb) \ve \btimes \blb - 5(\ve \bcdot \blb) \bl \btimes \blb + 2 \bl \btimes \ve \Big] + \cO(e^3),
\label{eq:dedtb}
\end{align}
where
\be
\omb(r) = \frac{3 G \Mb}{4 \ab^3 n}
\label{eq:omb}
\ee
characterizes the precession frequency of a disk annulus around the external binary.  Equations~\eqref{eq:dldtb} and~\eqref{eq:dedtb} include the effect of the quadrupole potential from the binary and are averaged over the binary period.

Internal hydrodynamical forces work to resist the differential nodal precession of the disk annuli, either in the form of bending waves \citep{PapaloizouLin(1995),LubowOgilvie(2000)} or viscosity \citep{PapaloizouPringle(1983),Ogilvie(1999)}, and enforce both coplanarity ($|\pd \bl/\pd \ln r| \ll 1$) and rigid body precession \citep{Larwood(1996), Xiang-GruessPapaloizou(2014)}.  Under their influence, the time evolution of the disk's unit angular momentum vector is given by
\begin{align}
\bigg( \frac{\pd \bl}{\pd t} \bigg)_{\rm int} + \bigg( \frac{\pd \bl}{\pd t} \bigg)_{\rm bin} &= \bomb (\bl \bcdot \blb) \bl \btimes \blb + \cO(e^2)
\label{eq:dldt} \\
\Rightarrow \bigg( \frac{\pd \bl}{\pd t} \bigg)_{\rm int} &= (\bomb - \omb) (\bl \bcdot \blb) \bl \btimes \blb + \cO(e^2),
\label{eq:dldti}
\end{align}
where $\bl$ is (nearly) independent of $r$, and
\be
\bomb = \frac{\int_{\ro}^{\rout} \Sg r^3 n \omb \der r}{\int_{\ro}^{\rout} \Sg r^3 n \der r}
\label{eq:bomb}
\ee
characterizes the precession frequency of the rigid disk around the binary.  The internal force that enforces rigid disk nodal precession must also act on $\ve$, so that $\ve$ remains perpendicular to $\bl$, i.e., 
\be
\left[ \frac{\pd (\ve \bcdot \bl)}{\pd t} \right]_{\rm int} = 0.
\ee
This requirement, together with the assumption that the internal force responsible for Eq.~\eqref{eq:dldti} is perpendicular to the disk, imply that the time evolution of the disk's eccentricity vector is
\be
\left( \frac{\pd \ve}{\pd t} \right)_{\rm int} = (\bomb - \omb) (\bl \bcdot \blb) \big[ \bl \bcdot (\ve \btimes \blb) \big] \bl + \cO(e^3).
\label{eq:dedti}
\ee
We justify Eq.~\eqref{eq:dedti} in the appendix.

Before we proceed, we comment on the validity of the assumption of coplanarity and rigid-body precession.  When the dimensionless Shakura-Sunyaev viscosity parameter $\ag$ satisfies $\ag \lesssim H/r$ ($H$ is the disk scaleheight), bending waves keep the disk coherent \citep{PapaloizouLin(1995),LubowOgilvie(2000)}. The amount of disk warp in this bending wave regime has been calculated in \cite{FoucartLai(2014)}, and assuming $p=1$ and $q=1/2$ [see Eqs. \eqref{eq:p}-\eqref{eq:q} in next section], is
\begin{align}
\bl&(\rout,t) - \bl(\rin,t) \approx
\nonumber \\
 &0.01\left( \frac{ \ag }{0.01} \right) \left( \frac{H(\rout)}{0.1 \, \rout} \right)^{-2} \left( \frac{ \Mb }{ M } \right) \left( \frac{ 3 \rout}{\ab} \right)^3 \frac{ \blb \btimes \bl(\rout,t) }{\sin I} 
 \nonumber \\
&- 0.01 \left( \frac{H(\rout)}{0.1 \, \rout} \right)^{-2} \left(\frac{\Mb}{M} \right)^2 \left( \frac{3 \rout}{\ab} \right)^6 \frac{[\blb \btimes \bl(\rout,t)] \btimes \blb}{ \sin I}.
\label{eq:dldlnr}
\end{align}
Numerical simulations give a similar result (e.g. \citealt{Larwood(1996), Xiang-GruessPapaloizou(2014), PicognaMarzari(2015)}).  On the other hand, when $\ag \gtrsim H/r$, viscous torques keep the disk coherent \citep{PapaloizouPringle(1983),Ogilvie(1999)}, and the disk diffusively damps to it's steady-state equilibrium warp profile over the timescale $t_{\rm visc} \sim 2 \ag r^2/(H^2 n)$ \citep{LodatoPringle(2007),LodatoPrice(2010),FoucartLai(2011)}.  Large warping and sometimes disk breaking is observed when the disk's viscous torque is comparable to or less than the torque exerted on the disk by the distant binary (e.g. \citealt{Larwood(1996),Dougan(2015)}).  Thus, the following derivation of the LK disk instability will be restricted to the $\alpha\lesssim H/r$ regime, which is applicable to protoplanetary disks.

For a flat disk, the effect of pressure on the time evolution of the disk's eccentricity is described by \citep{TeyssandierOgilvie(2016)}
\begin{align}
\bigg( &\frac{\pd \ve}{\pd t} \bigg)_{\rm press} = \bl \btimes \left[ \frac{1}{\Sg r^3 n} \frac{\pd}{\pd r} \left( \frac{\Sg \cs^2 r^3}{2} \frac{\pd \ve}{\pd r} \right) \right]
\nonumber \\
&+ \frac{1}{2 \Sg r n} \frac{\der( \Sg \cs^2)}{\der r} \bl \btimes \ve 
- \bl \btimes \left[ \frac{1}{2 \Sg r^3 n} \frac{\pd}{\pd r} \left( \Sg \frac{\der \cs^2}{\der r} r^3 \ve \right) \right] 
\nonumber \\
&+  \frac{3}{2 r^3 n} \frac{ \der(\cs^2 r^2)}{\der r} \bl \btimes \ve + \cO(e^2).
\label{eq:dedtp}
\end{align}
The last term in Eq. \eqref{eq:dedtp} arises from the disk's ``breathing mode," where the fluid displacements are proportional to $z^2$, where $z$ is the vertical coordinate of the disk \citep{Ogilvie(2008)}.  Earlier theories of eccentric disks do not include this term \citep{GoodchildOgilvie(2006)}.

Following \cite{TeyssandierOgilvie(2016)}, we also include the effect of bulk viscosity on the disk eccentricity evolution:
\be
\left( \frac{\pd \ve}{\pd t} \right)_{\rm visc} 
= \frac{1}{2 \Sg r^3 n} \frac{\pd}{\pd r} \left( \ag_{\rm b} \Sg \cs^2 r^3 \frac{\pd \ve}{\pd r} \right) + \cO(e^2),
\label{eq:dedtv}
\ee
small kinematic viscosity leads to over-stability, and a small bulk viscosity is needed to stabilize the eccentric disturbance \citep{Ogilvie(2001),LatterOgilvie(2006)}.

From Equation~\eqref{eq:dldt}, we see that the disk's unit angular momentum vector $\bl(t)$ precesses uniformly around $\blb$ with frequency $\om_{\rm prec} = -\bomb \cos I$, where $I$ is the inclination angle ($\cos I = \bl \bcdot \blb$).  Indeed, in the linear theory of LK oscillation of a test mass, the inclination stays constant while the eccentricity grows in time \citep{TremaineYavetz(2014)}.  To determine the stability of $\ve(r,t)$, it is necessary to consider the evolution equation of $\ve$ in the frame co-rotating with $\bl(t)$ \citep{TremaineYavetz(2014)}.  Including the gravitational perturbations and hydrodynamical effects, the time evolution of the disk's eccentricity vector $\ve$ is given by
\begin{align}
\bigg( \frac{\pd \ve}{\pd t} \bigg)_{\rm rot} =
&\left( \frac{\pd \ve}{\pd t} \right)_{\rm bin} + \left( \frac{\pd \ve}{\pd t} \right)_{\rm int} + \left( \frac{\pd \ve}{\pd t} \right)_{\rm press} 
\nonumber \\
&+ \left( \frac{\pd \ve}{\pd t} \right)_{\rm visc} + (\bomb \cos I) \blb \btimes \ve.
\label{eq:dedtrot}
\end{align}
We will work in this frame for the rest of the paper, and drop the subscript ``rot."

Define the complex eccentricity $E(r,t) \equiv \ve(r,t) \bcdot (\bx + \im \by)$, where $\by = \bl \btimes \blb/\sin I$ and $\bx = \by \btimes \bl$ are unit vectors, constant in the rotating frame.  Then Equation~\eqref{eq:dedtrot} becomes
\begin{align}
&\frac{\pd E}{\pd t} =  \im \omb \left[ 2 E - \frac{5 \sin^2 I}{2}(E + E^*) \right] 
\nonumber \\
&+ \im (\bomb - \omb) \cos^2 I E + \frac{\im}{\Sg r^3 n} \frac{\pd}{\pd r} \left( \frac{\Sg \cs^2 r^3}{2} \frac{\pd E}{\pd r} \right) 
\nonumber \\
&+ \frac{\im}{2 \Sg r n} \frac{\der (\Sg \cs^2)}{\der r} E - \frac{\im}{2 \Sg r^3 n} \frac{\pd}{\pd r} \left( \Sg \frac{\der \cs^2}{\der r} r^3 E \right)
\nonumber \\
 &+\frac{3 \im}{2 r^3 n} \frac{ \der(\cs^2 r^2)}{\der r} E + \frac{1}{2 \Sg r^3 n} \frac{\pd}{\pd r} \left( \agb \Sg \cs^2 r^3 \frac{\pd E}{\pd r} \right),
\label{eq:dEdt}
\end{align}
where $E^*$ denotes the complex conjugate to $E$.  To find the eigenmodes of Eq.~\eqref{eq:dEdt}, we separate $E$ into two ``polarizations":
\be
E(r,t) = E_+(r) \exp(\lam t) + E_-^*(r) \exp(\lam^* t).
\label{eq:Esol}
\ee
Here, $E_+$ and $ E_-$ are two complex functions, while $\lam$ is a complex eigenvalue.  Substituting Eq.~\eqref{eq:Esol} into Eq.~\eqref{eq:dEdt}, we obtain the coupled eigenvalue equations
\begin{align}
\lam &\Ep = \im \omb \left[ 2 \Ep - \frac{5 \sin^2 I}{2}(\Ep + \Em) \right] 
\nonumber \\
&+ \im (\bomb - \omb) \cos^2 I \Ep+ \frac{\im}{\Sg r^3 n} \frac{\der}{\der r} \left( \frac{\Sg \cs^2 r^3}{2} \frac{\der \Ep}{\der r} \right) 
\nonumber \\
&+ \frac{\im}{2 \Sg r n} \frac{\der (\Sg \cs^2)}{\der r} \Ep - \frac{\im}{2 \Sg r^3 n} \frac{\der}{\der r} \left( \Sg \frac{\der \cs^2}{\der r} r^3 \Ep \right) 
\nonumber \\
&+\frac{3 \im}{2 r^3 n} \frac{ \der(\cs^2 r^2)}{\der r} \Ep + \frac{1}{2 \Sg r^3 n} \frac{\der}{\der r} \left( \agb \Sg \cs^2 r^3 \frac{\der \Ep}{\der r} \right),
\label{eq:dEpdt} \\
\lam & \Em =  -\im \omb \left[ 2 \Em - \frac{5 \sin^2 I}{2}(\Ep + \Em) \right] 
\nonumber \\
&- \im (\bomb - \omb) \cos^2 I \Em- \frac{\im}{\Sg r^3 n} \frac{\der}{\der r} \left( \frac{\Sg \cs^2 r^3}{2} \frac{\der \Em}{\der r} \right) 
\nonumber \\
&- \frac{\im}{2 \Sg r n} \frac{\der (\Sg \cs^2)}{\der r} \Em + \frac{\im}{2 \Sg r^3 n} \frac{\der}{\der r} \left( \Sg \frac{\der \cs^2}{\der r} r^3 \Em \right) 
\nonumber \\
&-\frac{3 \im}{2 r^3 n} \frac{ \der(\cs^2 r^2)}{\der r} \Em + \frac{1}{2 \Sg r^3 n} \frac{\der}{\der r} \left( \agb \Sg \cs^2 r^3 \frac{\der \Em}{\der r} \right).
\label{eq:dEmdt}
\end{align}
When $\ag_{\rm b} = 0$, the eigenvalue $\lam$ is either real or imaginary.  Imaginary eigenvalues imply the eccentricity vector $\ve$ is precessing or librating around $\bl$, while real eignenvalues imply an exponentially growing or damping eccentricity.

For a thin ring ($\rin \simeq \rout $) 
of pressureless particles $(\cs = 0)$, Eqs.~\eqref{eq:dEpdt}-\eqref{eq:dEmdt} can be easily solved, giving
\be
\lambda^2 = - 2 \omb^2(2 - 5 \sin^2 I).
\ee
This recovers the standard results: eccentricity grows when $\ILK < I < 180^\circ - \ILK$ \citep{TremaineYavetz(2014)}, where
\be
 \ILK \equiv \sin^{-1} \sqrt{2/5} \simeq 39^\circ.
 \label{eq:ILK}
 \ee
 
\section{Results}
\label{sec:results}

To analyze the solutions of Eqs.~\eqref{eq:dEpdt} and~\eqref{eq:dEmdt}, we assume the disk surface density and sound-speed profiles of 
\be
\Sg(r) = \Sg(\rout) \left( \frac{\rout}{r} \right)^p
\label{eq:p}
\ee
and
\be
\cs(r) = \cs(\rout) \left( \frac{\rout}{r} \right)^q.
\label{eq:q} 
\ee
 A key dimensionless parameter in our analysis is the ratio
\begin{align}
S &\equiv \frac{\cs^2(\rout)}{\rout^2 n(\rout) \omb(\rout)}
\nonumber \\
&\simeq 0.36 \left( \frac{\ab}{3 \, \rout} \right)^3 \left( \frac{M}{\Mb} \right) \left( \frac{H(\rout)}{0.1 \, \rout} \right)^2,
\label{eq:S}
\end{align}
where we have approximated $\cs \simeq H n$ (where $H$ is the disk scale-height), and $\omb$ is defined in Eq.~\eqref{eq:omb}.  Physically, $S^{-1}$ measures the strength of the tidal torque (per unit mass) acting on the outer disc from the external companion $(r^2 n \omb)$ relative to the torque associated with gas pressure ($\cs^2$).

 Define the dimensionless radial coordinate $x \equiv r/\rout$, inner radius parameter $\xin \equiv \rin/\rout$, and dimensionless eigenvalue 
 \be
 \blam \equiv \lam/\omb(\rout).
 \label{eq:blam}
 \ee
 We assume that $\agb = \text{constant}$.  In terms of these parameters, Equations~\eqref{eq:dEpdt} and~\eqref{eq:dEmdt} become
\begin{align}
\blam & \Ep = \im x^{3/2} \left[ 2 \Ep - \frac{5 \sin^2 I}{2} (\Ep + \Em) \right]
\nonumber \\
&+ \im \left[ \frac{5/2 - p}{4 - p} \left( \frac{1-\xin^{4-p}}{1 - \xin^{5/2-p} } \right) - x^{3/2} \right] \cos^2 I \Ep
\nonumber \\
& + \im \frac{S x^{3/2 - 2q}}{2} \left[ \frac{\der^2}{\der x^2} + \left( \frac{3 - p}{x} \right) \frac{\der}{\der x} + \frac{A(p,q)}{x^2} \right] \Ep
\nonumber \\
&+ \ag_{\rm b} \frac{S x^{3/2 - 2q}}{2} \left[ \frac{\der^2}{\der x^2} + \left( \frac{3 - p - 2q}{x} \right) \frac{\der}{\der x} \right] \Ep,
\label{eq:dimdEpdt} \\
\blam & \Em = - \im x^{3/2} \left[ 2 \Em - \frac{5 \sin^2 I}{2} (\Ep + \Em) \right]
\nonumber \\
&- \im \left[ \frac{5/2 - p}{4 - p} \left( \frac{1 - \xin^{4-p}}{1 - \xin^{5/2 - p} } \right) - x^{3/2} \right] \cos^2 I \Em
\nonumber \\
&- \im  \frac{S x^{3/2 - 2q}}{2} \left[ \frac{\der^2}{\der x^2} + \left( \frac{3-p}{x} \right) \frac{\der}{\der x} + \frac{A(p,q)}{x^2} \right] \Em
\nonumber \\
& + \ag_{\rm b} \frac{S x^{3/2 - 2q}}{2} \left[ \frac{\der^2}{\der x^2} + \left( \frac{3-p-2q}{x} \right) \frac{\der}{\der x} \right] \Em,
\label{eq:dimdEmdt}
\end{align}
where\footnote
{If the breathing mode term is not included [last term in Eq.~\eqref{eq:dedtp}], $A(p,q) = 2q - p - 2 p q - 4 q^2$.  Equations \eqref{eq:dimdEpdt}-\eqref{eq:dimdEmdt} otherwise remain unchanged.}
\be
A(p,q) = 6 - 4 q - p - 2pq - 4q^2.
\label{eq:A}
\ee

We adopt a free boundary condition, where the eccentricity gradient vanishes on the disk's boundaries:
\be
\label{eq:bdry}
\left. \frac{\der E_\pm}{\der r} \right|_{r=\ro} = \left. \frac{\der E_\pm}{\der r} \right|_{r=\rout} = 0.
\ee

In the following subsections, we calculate the eigenvalues and eigenmodes to Eqs.~\eqref{eq:dimdEpdt} and~\eqref{eq:dimdEmdt}.  In Section~\ref{sec:thinannulus}, we investigate the limit $|\rout-\ro| \ll \rout$, where $\lam$, $\Ep(r)$, and $\Em(r)$ may be found analytically.  In Section~\ref{sec:extenddisk}, we calculate numerically $\lam$, $\Ep(r)$, and $\Em(r)$ for an inviscid ($\ag_{\rm b} = 0$) extended ($|\rout-\rin| \sim \rout$) disk.  In Section~\ref{sec:viscosity}, we investigate the effect of a non-zero bulk viscosity $\ag_{\rm b}$ on the eigenvalues $\lam$.

\begin{figure}
\centering
\includegraphics[scale=0.6]{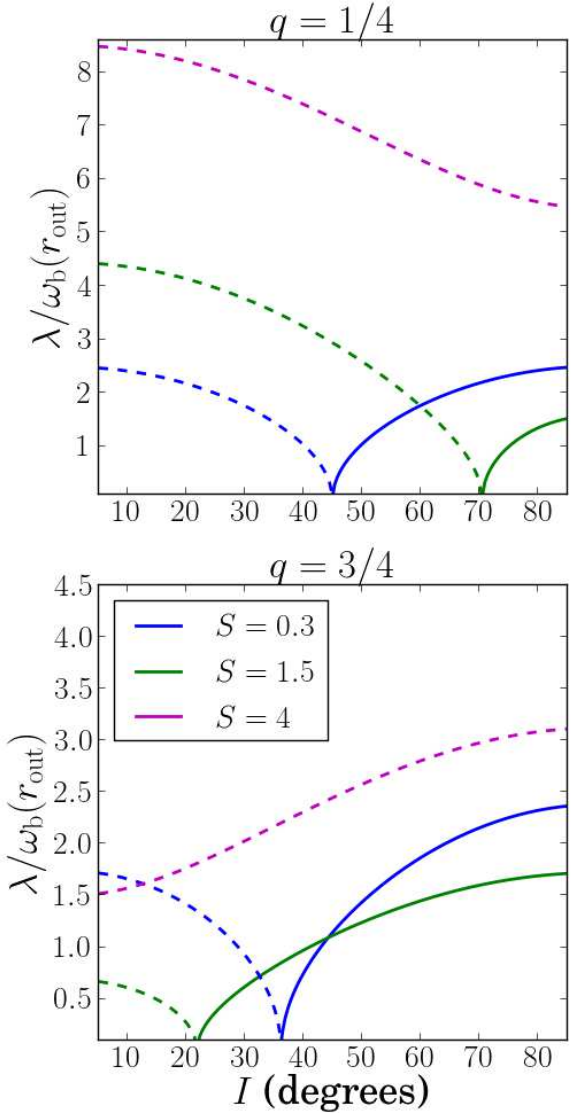}
\caption{
Real (solid) and imaginary (dotted) components of eigenvalue $\lam$ for a thin annulus [see Eqs.~\eqref{eq:lam_an} and~\eqref{eq:Esol}] as functions of inclination $I = \cos^{-1}(\bl \cdot \blb)$, for values of $S$ [Eq.~\eqref{eq:S}] and $q$ [Eq.~\eqref{eq:q}] as indicated.  We take $p = 1$ [Eq.~\eqref{eq:p}].}
\label{fig:lamI_an}
\end{figure}

\begin{figure}
\centering
\includegraphics[scale=0.6]{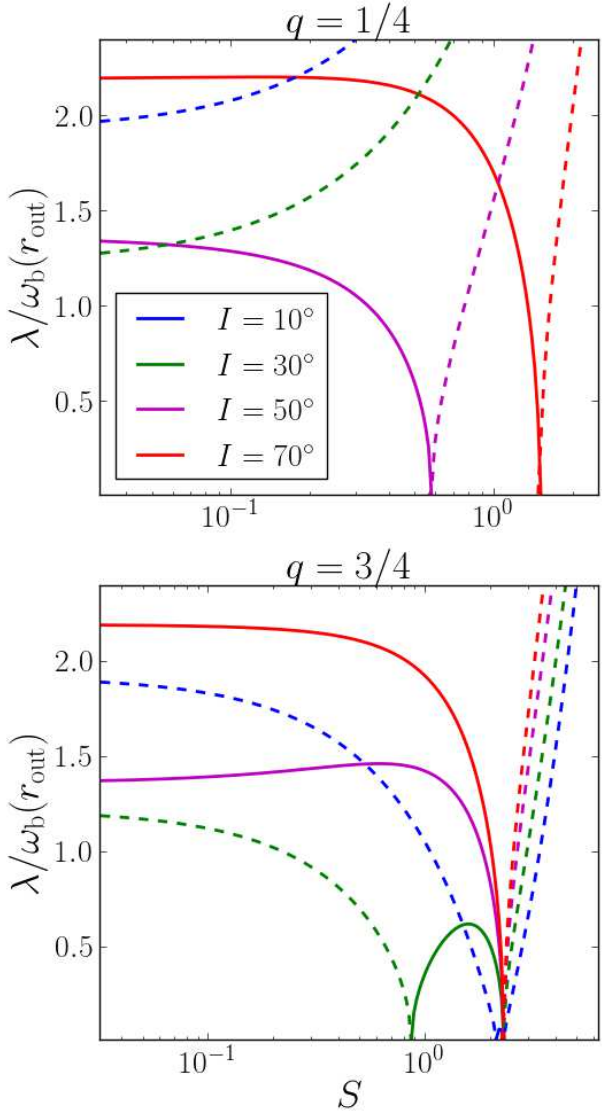}
\caption{
Real (solid) and imaginary (dotted) components of the eigenvalue $\lam$ [see Eqs.~\eqref{eq:lam_an} and~\eqref{eq:Esol}] as functions of $S$ [Eq.~\eqref{eq:S}], for values of inclination $I = \cos^{-1}(\bl \bcdot \blb)$ and $q$ [Eq.~\eqref{eq:q}] as indicated.  We take $p = 1$ [Eq.~\eqref{eq:p}].}
\label{fig:lamS_an}
\end{figure}

\subsection{Analytic Result for Thin Annulus}
\label{sec:thinannulus}

When $\rout-\ro \ll \rout$, we may expand all quantities in Equations~\eqref{eq:dEpdt} and~\eqref{eq:dEmdt} in terms of the small parameter $(\rout-r)/\rout=1-x$.  The boundary condition~\eqref{eq:bdry} and normalization condition~$\Ep(\rout) = 1$ imply
\be
\Ep(r) = 1 + \cO \left[ \left( 1-x \right)^3 \right]
\label{eq:Epconst}
\ee
and
\be
\Em(r) = \Em(\rout) + \cO \left[ \left(1-x \right)^3 \right].
\label{eq:Emconst}
\ee
Using the form of solutions~\eqref{eq:Epconst} and~\eqref{eq:Emconst}, we may solve for the eigenvalue $\blam$ [Eq.~\eqref{eq:blam}] to lowest order in $(\rout-\rin)/\rout = 1-\xin$:
\begin{align}
\blam^2 = \, - & \big[ 2 + S A(p,q)/2 \big] \big[(2-5 \sin^2 I) + S A(p,q)/2 \big].
\label{eq:lam_an}
\end{align}
The polynomial $A(p,q)$ is defined in Eq. \eqref{eq:A}, and $S$ in Eq.~\eqref{eq:S}.

Plotted in Figure~\ref{fig:lamI_an} are the real (solid) and imaginary (dashed) components of the eigenvalue $\lam$ given by Equation~\eqref{eq:lam_an}, as functions of inclination $I$ with values of $S$ as indicated.  We always show the solutions with ${\rm Re}(\lam) > 0$ and ${\rm Im}(\lam) > 0$.  When $S \ll 1$, we recover the classic LK result for a test particle, with $\lam^2 > 0$ when $I$ exceeds the critical inclination angle $\ILK$ [Eq. \eqref{eq:ILK}].  When $S \gg 1$, the Lidov-Kozai effect is suppressed by pressure gradients even when $I > \ILK$.  In general, the critical inclination angle for eccentricity growth increases with increasing $S$.  However, we see from Fig.~\ref{fig:lamI_an} that for $S = 1.5$ and $q=3/4$, the instability sets in when $I \gtrsim 22^\circ$.

Figure~\ref{fig:lamS_an} further illustrates the difference in behavior between $q = 1/4$ (top panel) and $q=3/4$ (bottom panel).  For $q = 1/4$, the real growth rate for inclinations $I > \ILK$ [Eq.~\eqref{eq:ILK}] monotonically decreases with increasing $S$, until $\lam$ becomes imaginary.  But for $q = 3/4$, a ``window of instability" opens for inclinations $I < \ILK$ when $S \sim 1$.

To understand the difference between these two models, consider the test particle limit ($\cs = 0$) and some additional pericenter precession $\domextra$ from a source other than the binary companion.  In the frame co-rotating with the test particle's orbit normal, the time evolution of the eccentricity vector is given by
\be
\frac{\der \ve}{\der t} = \omb \big[ 2 \bl \btimes \ve - 5 (\ve \bcdot \blb) \bl \btimes \blb \big] + \domextra \bl \btimes \ve.
\label{eq:TP}
\ee
Assuming $\ve \propto \exp(\lam t)$, we find the eigenvalue
\be
\lam^2 = -(2 \omb + \domextra)(2 \omb + \domextra - 5 \omb \sin^2 I).
\label{eq:lamTP}
\ee
When $\domextra \ge 0$, the extra pericenter precession works to suppress the LK instability, decreasing the range of $I$ values for eccentricity growth ($\lam^2 > 0$).  When $\domextra \le -2\omb$ or $\domextra \ge 3 \omb$, no value of $I$ is capable of exciting eccentricity growth.  But when $-2 \omb < \domextra < 0$, the extra precession works to cancel the pericenter precession induced on the test particle by the distant binary ($2 \omb$), thus increases the range of $I$ values for eccentricity growth.

Comparing Eq.~\eqref{eq:lamTP} to Eq.~\eqref{eq:lam_an} shows the pressure force in a disk annulus induces precession $\domextra = \omb S A(p,q)/2$.  Since $A(1,1/4)>0$, the pressure force in the $p = 1$ and $q=1/4$ disk tends to suppress eccentricity growth (Figs.~\ref{fig:lamI_an}-\ref{fig:lamS_an}, top).  But because $A(1,3/4)<0$, the pressure force in the $p = 1$ and $q=3/4$ disk can lead to eccentricity growth even when $I < \ILK$ (Figs.~\ref{fig:lamI_an}-\ref{fig:lamS_an}, bottom). 


\subsection{Inviscid Extended Disk}
\label{sec:extenddisk}

\begin{figure}
\centering
\includegraphics[scale=0.55]{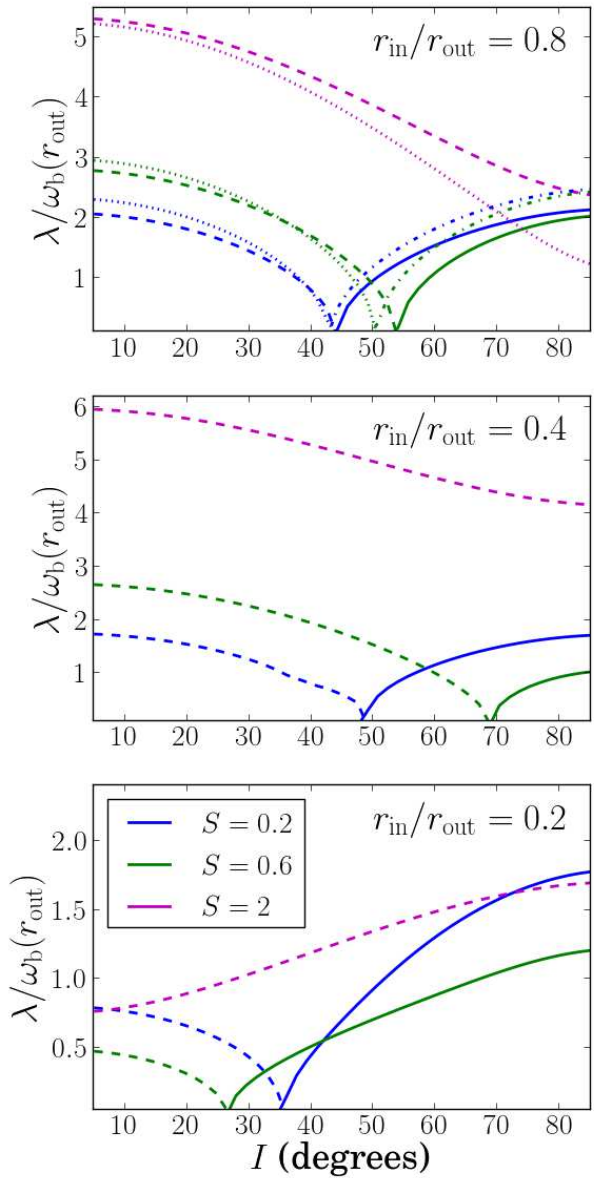}
\caption{
Real (solid lines) and imaginary (dashed lines) components of the eigenvalue $\lam$ for extended disks as a function of $I$.  We take $\ag_{\rm b} = 0$, $p=1$ [Eq.~\eqref{eq:p}], $q=1/4$ [Eq.~\eqref{eq:q}], with values of $S$ [Eq. \eqref{eq:S}] and $\rin/\rout$ as indicated.   In the top panel, we also plot the real (dot-dashed lines) and imaginary (dotted lines) components of the eigenvalue in the thin annulus limit [Eq. \eqref{eq:lam_an}].}
\label{fig:lam_rad}
\end{figure}

\begin{figure}
\centering
\includegraphics[scale=0.55]{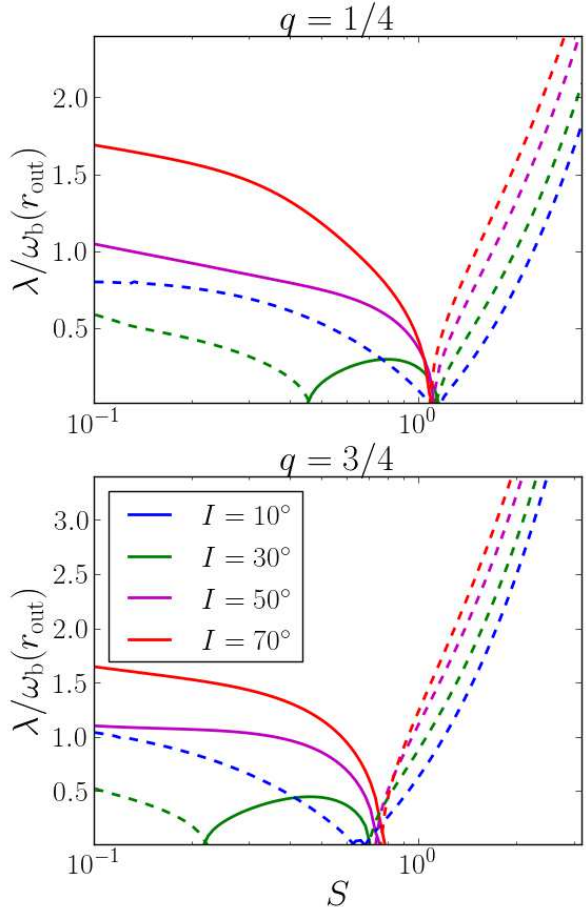}
\caption{
Real (solid lines) and imaginary (dashed lines) components of the eigenvalue for extended disks as a function of $S$.  We take $\ag_{\rm b} = 0$, $p=1$ [Eq. \eqref{eq:p}], $\rin/\rout = 0.2$, and values of $q$ [Eq.~\eqref{eq:q}] and inclination $I$ as indicated.}
\label{fig:lam_ext}
\end{figure}

We solve eigenvalue equations~\eqref{eq:dimdEpdt} and \eqref{eq:dimdEmdt} using the shooting method \citep{Press(2002)} for an inviscid ($\ag_{\rm b} = 0$)  extended ($|\rout-\rin| \sim \rout$) disk.  In Figure~\ref{fig:lam_rad}, we plot the real (solid) and imaginary (dashed) components of the eigenvalues $\lam = \blam \omb$ as functions of inclination $I$.  For $\rin/\rout$ close to unity, our numerical result agrees with the analytic expression for a thin annulus [Eq.~\eqref{eq:lam_an}].  In general, when $S \gg 1$, the pressure force suppresses the eccentricity growth for all values of $I$.  When $S \sim 1$, Fig.~\ref{fig:lam_rad} displays the importance of the disk's radial extent on the eigenvalues $\lam$.  For example, when $S = 0.6$ and $\xin = 0.4$, eccentricity growth is achieved for $I\gtrsim 69^\circ$, while for $\xin = 0.2$ the LK instability occurs for $I \gtrsim 27^\circ$.

In Figure \ref{fig:lam_ext}, we plot the eigenvalue $\lam = \blam \omb(\rout)$ as a function of $S$, for $\rin/\rout = 0.2$, $p=1$, and values of $q$ and $I$ as indicated.  Both models ($q = 1/4$ and $q = 3/4$) exhibit the suppression of eccentricity growth for $S \gtrsim 1$, and both models have a window of instability open when $S \sim (\text{few}) \times 0.1$.  This window of instability is similar to that seen in Figure~\ref{fig:lamS_an}.


\begin{figure}
\centering
\includegraphics[scale=0.65]{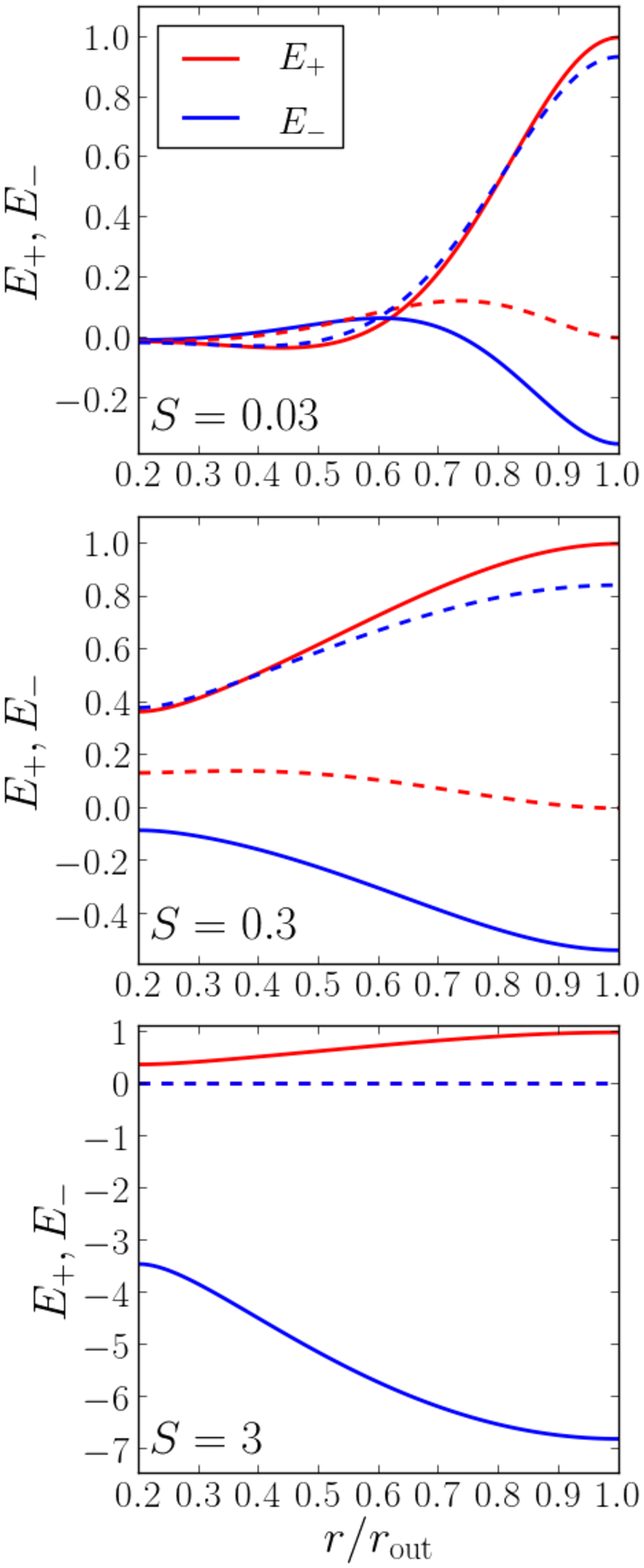}
\caption{
Real (solid lines) and imaginary (dashed lines) components of the eigenfunctions $\Ep(r)$ and $\Em(r)$ for an extended disk.  The normalization condition is $\Ep(\rout) = 1$.  The disk parameters are $\ag_{\rm b} = 0$, $p=1$, $q=3/4$, $\rin/\rout = 0.2$, inclination $I = 70^\circ$, and values of $S$ [Eq.~\eqref{eq:S}] as indicated.   The corresponding eigenvalues are $\blam = 1.84$ (top), $\blam = 1.44$ (middle), and $\blam = 5.57 \im$ (bottom).}
\label{fig:E}
\end{figure}

Figure~\ref{fig:E} depicts some examples of the eigenfunctions $E_+(r)$ and $E_-(r)$ for disk models with $S = 0.03, 0.3$, and $3$.  We see that for small $S$ (top panel), the amplitudes $|E_+|$ and $|E_-|$ are largest at $r = \rout$ and decreases rapidly as $r \to \rin$.  For larger $S$ (middle and lower panels), the variations of $|E_+|$ and $|E_-|$ across the disk become smaller as the larger sound speed ``smooths out" the disk.  The bottom panel of Fig.~\ref{fig:E} shows that when $S = 3$ (for which the disk is stable since $\lam$ is imaginary), the eigenfunctions $\Ep$ and $\Em$ are both real and satisfy $\Em > \Ep$, implying retrograde precession of the disk's eccentricity.

\subsection{Effect of Viscosity}
\label{sec:viscosity}

\begin{figure}
\centering
\includegraphics[scale=0.55]{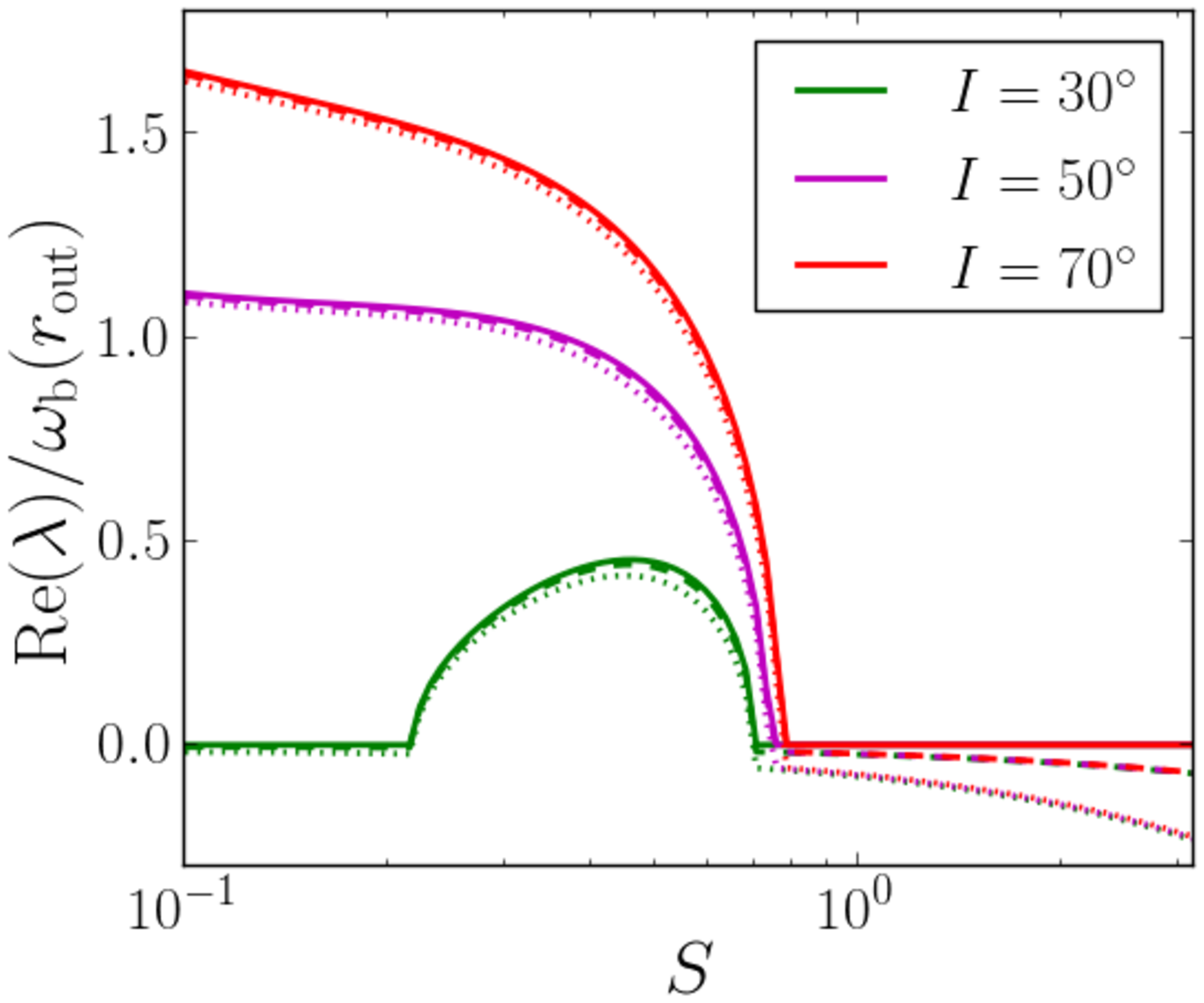}
\caption{
Real parts of the eigenvalues $\lam$ for extended disks, plotted as a function of $S$, with $\ag_{\rm b} = 0$ (solid), $\ag_{\rm b} = 0.03$ (dashed), and $\ag_{\rm b} = 0.1$ (dotted).  The other disk parameters are $p = 1$, $q = 3/4$, $\rin/\rout = 0.2$, and the values of $I$ are as indicated.}
\label{fig:lam_visc}
\end{figure}

We solve the eigenvalue equations~\eqref{eq:dEpdt}-\eqref{eq:dEmdt} including the viscosity term.  In Figure \ref{fig:lam_visc}, we plot the real parts of the eigenvalues $\lam$ for $\ag_{\rm b} = 0, 0.03$, and $0.1$.  When $S \lesssim 1$, we see for a range of inclinations, the growth rates are only slightly modified by viscosity.  When $S \gtrsim 1$, the addition of a small viscosity begins to be important.  However, in this regime, the instability is already suppressed by the disk's pressure, so the additional damping from $\ag_{\rm b}$ when $S \gtrsim 1$ is not relevant for the LK effect.   We conclude that a small bulk viscosity does little to quench the LK instability.


\section{Summary and Discussion}
\label{sec:discussion}

\subsection{Summary of Key Results}
\label{sec:summary}

Using linear theory of eccentric disturbances in hydrodynamical disks,
we have shown that circumstellar disks in binary systems may undergo
coherent eccentricity growth when the disk is significantly inclined
with respect to binary orbital plane. We consider the regime where the
disk remains approximately flat and undergoes rigid-body nodal
precession around the binary; this requires that bending waves
efficiently communicate warps in different regions of the disk within
the precession period. We find that the disk's eccentricity response to the 
secular tidal forcing from the binary companion depends crucially on the dimensionless
ratio [see Eq.~\eqref{eq:S}], 
\be
S=\left( \frac{\cs^2}{ 3G \Mb r^2/4\ab^3}\right)_{r=\rout},
\ee
where $\cs^2$ (disk sound speed squared) measures the characteristic
torque (per unit mass) associated with gas pressure, 
$3G\Mb r^2/4\ab^3$ (with $\Mb$ and $\ab$ the companion mass and semi-major
axis) measures the tidal torque from the companion. 
The eccentricity response also depends on the disk's radial extent 
($r_{\rm out}/r_{\rm in}$) and density and sound speed profiles [Eqs.~\eqref{eq:p} and~\eqref{eq:q}].
\begin{enumerate}
\item When $S \ll 1$, the ``standard" Lidov-Kozai effect is reproduced for a thin
disk annulus ($r_{\rm out}/r_{\rm in}\rightarrow 1$), with exponential
eccentricity growth occuring for disk inclination $I$ (with
respect to the binary orbital plane) between $39^\circ$ and $141^\circ$.
\item As $S$ increases, the inclination window for disk eccentricity growth generally 
decreases. When $S \gg 1$, eccentricity growth is completely quenched for all disk inclinations.
\item When $S \sim 1$, a new ``window of instability" opens up for certain disk parameters,  
where coherent disk eccentricity growth is observed for inclinations $I$ outside the
standard $(39^\circ,141^\circ)$ window.
\end{enumerate}
These conclusions are qualitatively robust, shown through both analytic
calculations when the disk's radial extent is negligible (thin annulus; 
Sec.~\ref{sec:thinannulus}) and numerical eigenmode analyses when the disk
has a significant radial extent (Sec.~\ref{sec:extenddisk}).
We find that viscosity does little to quench the Lidov-Kozai instability of the disk
(Sec.~\ref{sec:viscosity}).

The different disk eccentricity responses to the secular tidal forcing 
can be understood in terms of the apsidal precession produced by gas pressure
(i.e. \citealt{Papaloizou(2002),GoodchildOgilvie(2006),TeyssandierOgilvie(2016)}).
This precession depends on the $S$ and the disk density/pressure
profiles.  Unlike the other short-range forces, such as those due to
General Relativity and tidal interaction in hierarchical triple
systems (e.g. Liu et al.~2015), the pressure-induced precession can be
either prograde or retrograde, depending on the disk profiles
[see Eq.~\eqref{eq:lam_an}; see also \citealt{TeyssandierOgilvie(2016)}].  This
gives rise to the nontrivial behavior of the disk's eccentricity
response for $S\sim 1$.

\subsection{Discussion}
\label{sec:discuss}

In this paper we have focused on the linear regime of the disk
Lidov-Kozai instability, which manifests as the coherent growth of
disk eccentricity, with no change in the disk inclination (which
enters at the order $e^2$). Numerical simulations are necessary to
fully understand the nonlinear development of the disk
eccentricity-inclination oscillations \citep{Martin(2014),Fu(2015a),Fu(2015b)}.
Nevertheless, our analytic results can be used to determine under what
conditions a hydrodynamical circumstellar disk is susceptible to
Lidov-Kozai oscillations, without resorting to full 3D numerical
simulations.

We note that the dynamical behavior of eccentric disturbances in a
hydrodynamical disk depends on the disk's equation of state and
vertical structure \citep{GoodchildOgilvie(2006),Ogilvie(2008),TeyssandierOgilvie(2016)}.
We have adopted the eccentric disk models with locally isothermal
equation of state, including the 3D breathing mode term from the disk's vertical structure (see \citealt{Ogilvie(2008)} for discussion). Using different models can change the details of
our results, but not the general conclusions summarized in Section~\ref{sec:summary}.

The disk eccentricity excitation mechanism studied in this paper is
distinct from the mechanism that relies on eccentric Lindblad
resonance \citep{Lubow(1991)}. The latter operates on the dynamical timescale
and requires that the disk be sufficiently extended relative to the
binary separation (i.e., $r_{\rm out}/a$ is sufficiently larger) so
that the resonance resides in the disk. By contrast, the disk Lidov-Kozai 
mechanism for eccentricity excitation requires an inclined binary companion, and
operates on a secular timescale [Eq.~\eqref{eq:omb}]
\begin{eqnarray}
&& t_{\rm LK}\sim \omb(r_{\rm out})^{-1}
= 5.7\times 10^3\,{\rm years}\left(\frac{M}{\Mb}\right)\left(\frac{\ab}{3 \rout}\right)^3 
\nonumber\\
&&\times\left(\frac{M}{1 M_\odot}\right)^{-1/2}\left(\frac{\rout}{100 \,\text{AU}}\right)^{3/2}.
\end{eqnarray}
For protoplanetary disks, this timescale is much less than the disk lifetime (a few Myrs).
To avoid suppression of the instability by the gas pressure, we also require
\begin{equation}
S = 0.36 \left( \frac{\ab}{3\,\rout}\right)^3 \left(\frac{M}{\Mb}\right)
\left( \frac{H(\rout)}{0.1 \, \rout} \right)^2 \lesssim 1.
\label{eq:condS}
\end{equation}
Thus, a ``weaker'' companion (large $\ab$ and small $\Mb$) would not excite
eccentricity in a thick (large $H/R$) disk.  Condition~\eqref{eq:condS} is consistent with the SPH simulations of \cite{Martin(2014)} and \cite{Fu(2015a)}, where $S$ values in the range $8.5\times10^{-3}$ to $0.11$ were used.

Finally, for a massive disk, the LK instability can be suppressed due to apsidal precession generated by disk self-gravity \citep{Batygin(2011),Fu(2015b)}. The apsidal precession rate from the disk's self gravity is roughly
\be
\om_{\rm sg}(r) \sim \frac{\pi G \Sg}{r n}.
\ee
Crudely, to avoid suppression of the LK instability, we require $\om_{\rm sg}(\rout) \lesssim \omb(\rout)$, or the disk mass
\be
\Md \lesssim \Mb \left(\frac{\rout}{\ab}\right)^3 \sim 0.04 \, \Mb \left( \frac{3 \rout}{\ab} \right)^3.
\ee

\section*{Acknowledgments}

We thank the anonymous referee for his or her valuable comments.  This work has been supported in part by NASA grants NNX14AG94G and
NNX14AP31G, and a Simons Fellowship from the Simons Foundation.  JZ is
supported by a NASA Earth and Space Sciences Fellowship in
Astrophysics.

\section*{Appendix}

 This appendix is devoted to the derivation of Eqs.~\eqref{eq:dldti} and~\eqref{eq:dedti}.  Our key assumption is that the internal force in the disk acts to enforce coplanarity and rigid body precession of the disk.
 
 Consider a disk particle (test mass) with the position vector $\br$ and velocity $\bv$ relative to the central star. It's angular momentum is $\bL = \br \btimes \bv$, and its eccentricity vector is
 \be
 \ve = \frac{1}{G M} \bv \btimes (\br \btimes \bv) - \frac{\br}{r}.
 \ee
 Under the action of a perturbing force $\bF$, the vectors $\bL$ and $\ve$ evolve according to
 \begin{align}
 \frac{\pd \bL}{\pd t} &= \br \btimes \bF,
 \label{eq:dLdt} \\
 \frac{\pd \ve}{\pd t} &= \frac{1}{GM} \bF \btimes (\br \btimes \bv) + \frac{1}{GM} \bv \btimes (\br \btimes \bF).
 \label{eq:dvedt}
 \end{align}
The perturbing force $\bF = \bF_{\rm b} + \bF_{\rm int}$ consists of the tidal force from the binary companion $\bF_{\rm b}$ and the internal pressure force $\bF_{\rm int}$.  To quadrapole order, the tidal force is given by
\be
\bF_{\rm b} = \frac{G \Mb}{|\brb|^3} \left[ \br - 3 \frac{\brb (\br \bcdot \brb)}{|\brb|^2} \right],
\ee
where $\Mb$ and $\brb$ are the mass and position vectors of the companion.  Take the binary to be on a circular orbit with semi-major axis $\ab$ and mean anomaly $\phi_{\rm b}$, and let $\vr$, $\vphi = \bl \btimes \vr$, and $\bl$ be the radial, azimuthal, and angular momentum unit vectors of the test mass, respectively.  Averaging over the binary's orbital motion, we obtain the averaged tidal force
\begin{align}
\bar \bF_{\rm b} \equiv &\frac{1}{2\pi} \int_{0}^{2\pi} \bF_{\rm b} \der \phi_{\rm b}
\label{eq:avedef}\\
= & \frac{2}{3} r n \omb \big( 1 - 3 \sin^2\varphi \sin^2 I \big) \hr
\nonumber \\
&- 2 r n \omb  \big( \sin\varphi \cos\varphi \sin^2 I \big) \hphi
\nonumber \\
&- 2 r n \omb \big( \sin \varphi \sin I \cos I \big) \bl,
\label{eq:fbave}
\end{align}
where $\omb$ is defined in Eq.~\eqref{eq:omb}, and $\varphi = \om + f$ is the azimuthal angle of the test mass measured from the ascending node ($\om$ and $f$ are the argument of pericenter and true anomaly).  The $\vr$ and $\vphi$ components of $\bar \bF_{\rm b}$ do not change $\bL$, and the $\bl$ component induces precession at a rate $-\omb \cos I \blb$ [see Eq.~\eqref{eq:dldtb}].  To ensure coplanarity and rigid-body precession of test particles at different radii, we assume that the internal force from disk pressure has the form
\be
\bF_{\rm int} = -2 r n (\bomb - \omb) \big(\sin \varphi \sin I \cos I \big) \bl,
\label{eq:fint}
\ee
where $\bomb$ is given in Eq.~\eqref{eq:bomb}.

We now substitute Eq.~\eqref{eq:fint} into Eqs.~\eqref{eq:dLdt} and~\eqref{eq:dvedt} to obtain the effect of $\bF_{\rm int}$ on $\bl$ and $\ve$.  For a disk particle on an eccentric orbit $e \ll 1$, we can expand $r$ and $f$ in powers of $e$ \citep{MurrayDermott(1999)}.  Averaging over the mean anomaly of the test particle, we obtain Eqs.~\eqref{eq:dldti} and~\eqref{eq:dedti}.

\end{document}